\documentstyle[preprint,aps,,epsf]{revtex}
\begin{document}

\draft

\title{Mesoscopic mechanism of exchange interaction in magnetic
multilayers}

\author{A. Zyuzin}

\address{Max-Planck-Institut f$\ddot u$r Physik Komplexer Systeme,
01187 Dresden Germany }
\address
{A.F.Ioffe Physical- Technical Institute, 194021, St.Petersburg,
Russia}

\author{B. Spivak}

\address{Physics Department, University of Washington, Seattle, WA 98195, 
USA}
\address{  Max-Planck-Institut f$\ddot u$r Festkorperforschung Hochfeld-Magnetolabor,
Grenoble, France}

\author{I. Vagner, P.Wyder}

\address{  
Max-Planck-Institut f$\ddot u$r Festkorperforschung Hochfeld-Magnetolabor,
Grenoble, France}

\maketitle

\begin{abstract}
We discuss a mesoscopic mechanism of exchange interaction in ferromagnet-normal
metal-ferromagnet multilayers. We show that in the case when the metal's
thickness is larger than the electron mean free path, the relative
orientation of magnetizations in the ferromagnets is perpendicular.
The exchange energy between ferromagnets decays with the metal thickness as
a power law. 
\end{abstract}


\newpage

Both the experiment and the theory of ferromagnet-normal
metal-ferromagnet multilayers have attracted a lot of attention
$^{[1-6]}$.  
An example of such a structure consisting of two ferromagnetic
films separated by a nonferromagnetic metallic film is shown in Fig.1.
In the case when the metal thickness $L$ is
much smaller
than 
the electron scattering mean free path $l$  
the sign of the exchange interaction
energy
between the ferromagnet's magnetizations oscillates  as a
functions of $L$ with a period of order of the Fermi wave length.
As a result the magnetic structure of the system
oscillates between ferromagnetic and antiferromagnetic 
orientations of the ferromagnet's magnetizations $^{[1-6]}$.
The explanation of this phenomenon is based on the fact that the interlayer 
exchange energy is due to Ruderman-Kittel interaction between
electron spins in different ferromagnets.
 
In the case of low temperatures and at $|\bbox{r}-\bbox{r}'|\gg l$ the
exchange
 energy
between two localized spins $<J(\bbox{r},\bbox{r}')>$ averaged over the
scattering potential
configurations decays exponentially
 with $|\bbox{r}-\bbox{r}'|$ $^{[7]}$. Here $\bbox{r}$ and $\bbox{r}'$ are
coordinates of spins and brakets $< >$ stand for averaging over realizations of
 the scattering
potential in the metal and the ferromagnets.
Recent experiments on ferromagnet-metal-ferromagnet
 multilayers
  $^{[8]}$   
 imply, however, that the exchange energy between the ferromagnets does
not decay exponentially at $L\gg l$
and that the equilibrium relative orientation of the ferromagnet's
magnetizations is
perpendicular independently of $L$.  
Phenomenologically, this 
 situation can be described by an effective energy per unit area
 \begin{equation}E=-J^{0}(\bbox{m}^{0}_{1}\cdot\bbox{m}^{0}_{2}
)+B[(\bbox{m}^{0}_{1}
\cdot\bbox{m}^{0}_{2} 
)^{2}
-1]
\end{equation}
in the case $2B \gg | J^{0} |$.
Here $\bbox{m}^{0}_{1} $ and $\bbox{m}^{0}_{2}$ are, averaged over the
film's volumes, unit vectors parallel to 
magnetizations of the
ferromagnetic films. Indices 1,2 indicate
 the first and the second
ferromagnetic film respectively, $J^{0}$ and $B(\theta)$ are bilinear and
biquadratic
coupling coefficients. In general, $B(\theta)$ is a smooth function
of
the angle $\theta$ between $\bbox{m}^{0}_{1} $ and $\bbox{m}^{0}_{2}$. 

In this paper we discuss a theory of this phenomenon.
It has been shown in $^{[9-11]}$ that the exponential decays of the average
$<J(\bbox{r},\bbox{r}')>$ is connected to the fact that it
has a random
sign at large $L$. The modulus of the exchange
interaction decay with $L$ as a power law.
 
 We can introduce a local exchange energy $J (\bbox{\rho})$ between the
ferromagnets as an average
of
$J(\bbox{r},\bbox{r}')$ over a ferromagnet's surface area of order of
$L^2$.
Here $\bbox{\rho}$
is coordinate along the films. We assume that $J (\bbox{\rho})$ is small
enough and spatial dependence of the magnetizations on the scale of
order of $L$ can be neglected.
 According to Slonczewski $^{[12,13]}$, biquadratic term proportional to
 $B$ in Eq.1
 can originate from the existence of spatial fluctuations of the sign of
exchange
interaction
 $J (\bbox{\rho}) = \langle J\rangle + \delta J (\bbox{\rho})$ along the
layers.
The fluctuations of
$J(\bbox{\rho})$ cause fluctuations of
directions
of magnetizations.
Energy associated with spatial fluctuations in
the magnetization's directions can be represented as
\begin{eqnarray}
E(J(\bbox{\rho}) , \bbox{m}_i (\bbox{\rho}))= -\int d^2 \bbox{\rho} 
J(\bbox{\rho})(\bbox{m}_{1}(\bbox{\rho})\cdot\bbox{m}_{2} (\bbox{\rho})) 
\nonumber \\
 +  \alpha d \int d^2
\bbox{\rho} 
\biggl[\frac {\partial\bbox{m}_1
(\bbox{\rho} )}
{\partial\bbox{\rho}} \cdot \frac {\partial\bbox{m}_1 (\bbox{\rho}
)}{\partial\bbox{\rho}}+
\frac
{\partial\bbox{m}_2 (\bbox{\rho} )}{\partial\bbox{\rho}}
\cdot \frac{\partial\bbox{m}_2 (\bbox{\rho}
)}{\partial\bbox{\rho}}\biggr]
\end{eqnarray}
where the first term corresponds to the interfilms exchange energy,
the second term is associated with the gradients of magnetizations
inside the films, $d$ is the ferromagnetic film's thickness and
$\alpha$ is a coefficient characterizing the exchange energy value in the
ferromagnets.

In the case when $\delta J(\bbox{\rho})\gg \langle J\rangle $ and
$J(\bbox{\rho})$ has
a random sign, the
energy
$E(J , \bbox{m}_i (\bbox{\rho}))$ has a minimum at a sample specific
 realization  
$\bbox{m}_i (\bbox{\rho} )=\bbox{m}^{0}_{i} + \delta \bbox{m}_i (\bbox{\rho} ; [\delta J])
$ with $\bbox{m}^{0}_{1} \bot \bbox{m}^{0}_{2}$ $^{[12,14]}$ and 
\begin{eqnarray} 
B \equiv  \frac {B_0}{\alpha d} G(\theta) \\
G= \int d^2 {\bbox{\rho}} \langle \delta J (\bbox{\rho}) \delta
J(0)\rangle 
\end{eqnarray} 
Here
$B_0$
is a
number of order unity
$^{[14]}$.

Let us consider the case when $J(\bbox{\rho})$
has random sign due to mesoscopic fluctuations of Ruderman-Kittel
oscillations inside the metal $^{[9-11]}$.
We assume that the ferromagnetic film's thickness $d>L_s=\sqrt
{D/\omega_s}$ and 
 $L_T=\sqrt{\frac{D}{T}} > d, L$ . The latter inequality allows us to neglect
 temperature
dependence of $B$. Here $D$ is the diffusion constant, which is assumed to be
 the same in the
ferromagnetic and nonferromagnetic parts of the sample, and $\omega_{s}$ is
the exchange 
spin splitting energy in the ferromagnets. 
We will show that in the case $L_{s}<L$,
\begin{equation}
 G = \gamma (\theta)(\frac {E_{c}}{L})^{2}  
\end{equation}
while in the case $L_{s}>L$,
\begin{equation}
 G = \gamma_1 (\theta)(\frac {\omega_{s}}{L_s})^{2}.  
\end{equation}
Here $\gamma$ and $\gamma_{1}$ are smooth functions of $\theta$ of order
 unity and $E_{c}=\frac{D}{L^{2}}$ is the Thouless energy.
Qualitatively, Eqs.5,6 can be understood as follows:
In the case $|\bbox{r}-\bbox{r}'|\gg l $ the random oscillations of
$J(\bbox{r},\bbox{r}')$ exhibit a long range sign correlations $^{[15]}$.
In the case $L_{s}\ll L$ these long range correlations should be cut of at
a length of the order of $L$. As a result, the fluctuations of the exchange
energy averaged over the area of order of $L^{2}$ is of order
$E_{c}$; and they are $\delta$-correlated at a
distances larger than $L$. This leads to Eq.5. 
We think that the estimate presented above can be relevant
for the experiment $^{[8]}$.
In the opposite limit $L_{s}\gg L$ the cut off length is $L_{s}$.  The
fluctuations of the exchange energy   
averaged over the area of order $L^{2}_{s}$ is of order $\omega_{s}$.
This leads to Eq.6, which is independent of $L$.

To derive the results presented above we describe the exchange energy
splitting
in
ferromagnet with the help of an effective  Hamiltonian 
\begin{equation} 
H = H_0 + \bbox{h}(\bbox{r}; \theta) \bbox{\sigma} 
\end{equation}
Here $H_0$ is the Hamiltonian of free electron gas in a random potential
 $U(\bbox{r})$, $\bbox{h}(\bbox{r},\theta) \equiv \omega_s
\bbox{m}(\bbox{r},\theta )$ is the effective magnetic filed which is acting
only on electron spins, $\bbox{m} (\bbox{r}, \theta )$ denotes a unit
vector parollel to the magnetic moment
in ferromagnets in the case when the angle between $\bbox{m}_{1}^{0}$ and
$\bbox{m}_{1}^{0}$  is
 $\theta$, and $\bbox{\sigma}=\{\sigma_{x},\sigma_{y},\sigma_{z}\}$ is
the Pauli matrixes vector.
 We
assume the following correlation properties of
random potential:
 $<U(\bbox{r})>=0$
and $<U(\bbox{r})U(\bbox{r}')>=\frac {1}{2\pi \nu_0
\tau}\delta (\bbox{r} - \bbox{r}')$.
Here $\nu_0$ is density of states at fermi level, $\tau$ is mean free
 scattering time  
of electrons.

To get the correlation function $\langle \delta J(0)\delta
J(\bbox{\rho})\rangle$ we
consider sample specific fluctuations of
thermodynamic potential $\Omega (\theta)$ of the electrons as a function of
$\theta$,
\begin{equation}
\Omega (\theta) = \langle \Omega (\theta)\rangle + 
\delta \Omega (\theta).
\end{equation}
Using the identity $\frac {d\Delta\Omega (\theta)}{d\theta}= \int d^2\bbox{\rho}
\delta J (\bbox{\rho})$ we get
\begin{equation} 
G(\theta)= \langle (\frac {d \delta \Omega }{d \theta} )^2 \rangle 
\end{equation}  
In the case of noninteracting electrons we can 
express thermodynamic potential as $\Omega = \int\limits_{0}^{\mu} d\mu N(\mu)$ ,
where $N(\mu)$ is the number of electrons at given chemical potential
$\mu$.
Then, the correlation function of fluctuations of thermodynamic
potential 
has the form
\begin{equation} 
\langle \delta \Omega (\theta _1) \delta \Omega (\theta _2)\rangle =
\int \limits_{0}^{\mu }
d\mu _1 d\mu _2 \langle \delta N(\mu_1 ,\theta_1)\delta N(\mu_2 , \theta_2)
 \rangle 
\end{equation}
To calculate it we use the usual diagram technique for averaging over
configurations of
disordered potential
$^{[16]}$.
Diagrams for correlation function of number 
of electrons are shown in Fig.2.
As a result we have
\begin{eqnarray}
 \langle \delta N (\mu_1 ,\theta_1) 
\delta N (\mu_2 , \theta_2)  \rangle = \frac{2}{\pi} 
T \sum_{\omega_n >0} \omega_n Re\int_{V} d^3 \bbox{r} d^3 \bbox{r'}
D_{\alpha\beta}^{\gamma\nu} (\bbox{r} , \bbox{r'} ; \omega_n )
D_{\beta\alpha}^{\nu\gamma} (\bbox{r'} , \bbox{r} ; \omega_n)=
\nonumber\\  =\frac{2}{\pi} 
T \sum_{\omega_n >0} Re \int \frac {d^2 \bbox{q}}{(2\pi )^2} \sum_{m}
 \frac {\omega_n}{(E_m (\theta _1 ,\theta _2 ) + D\bbox{q}^2 +
\omega_n + i(\mu _1 -\mu _2 ) )^2} 
\end{eqnarray}
where $\omega_n = 2\pi n T$ is the Matsubara frequency, $n=1,2..$ and
$\alpha, \beta, \gamma, \mu$ are spin indices.
Diffusion propagators 
$D_{\alpha\beta}^{\gamma\nu} (\bbox{r}, \bbox{r}' ; \omega_n )$
obey the equation
\begin{eqnarray}
\biggl[ [- D\triangle +\omega_n +i(\mu _1 -\mu _2 )]\delta_{\alpha\xi}
\delta_{\gamma\mu}
+ i(\bbox{h}(\bbox{r} ;\theta_1) \bbox{\sigma}_{\gamma \mu}\delta_{\alpha\xi} - 
\bbox{h}(\bbox{r} ;\theta_2)
\bbox{\sigma}_{\alpha\xi}\delta_{\gamma\mu}) \biggr]
D_{\xi\beta}^{\mu\nu} (\bbox{r} , \bbox{r'} ; \omega_n )= 
\nonumber\\= 
\delta (\bbox{r} - \bbox{r'} ) \delta _{\gamma ,\nu} \delta_{\alpha , \beta}
\end{eqnarray}
The second equality in Eq.11 is the representation in terms of
 eigenvalues of Eq.12.
 In the case of geometry of
the system, shown in Fig.1, the eigenvalues are equal to 
$D\bbox{q}^2  + E_m (\theta _1 ,\theta _2 )$. Here the 
spectrum $E_m (\theta _1 ,\theta _2 )$ is determined by equation
\begin{equation} [- D\delta_{\alpha\xi}
\delta_{\gamma\mu}\frac {d^2}{dz^2} +i(\bbox{h}(\bbox{r} ;\theta_1) \bbox{\sigma}_{\gamma \mu}\delta_{\alpha\xi} - 
\bbox{h}(\bbox{r} ;\theta_2)
\bbox{\sigma}_{\alpha\xi}\delta_{\gamma\mu})]\Psi _m (z; \mu , \xi) = E_m \Psi _m (z;\gamma , \alpha) 
\end{equation}

To calculate Eq.11 we
 use following equalities
\begin{eqnarray}  
\int \frac {d^2 \bbox{q}}{(2\pi )^2}   
 \sum_{m} 
(E_m + D\bbox{q}^2 +\omega_n + i(\mu _1 -\mu _2 ) )^{-2} 
= \nonumber\\
=-\frac {d}{d\omega_n}\int \frac {d^2 \bbox{q}}{(2\pi )^2}  \frac {1}{4\pi i}
\int_{c} dp\frac {1}{\frac {Dp^2}{L^2}+D\bbox{q}^2 +
\omega_n + i(\mu _1 -\mu _2 )} \frac {d\ln det(p)}{dp}=
\nonumber\\
=\frac {1}{8\pi D} \frac {d}{d\omega_n}\ln (det(ip_0) det(-ip_0))
\end{eqnarray}
Here $det(p)=\prod\limits_{m}(p-E_m) $ is the spectral determinant of
Eq.13, and
$p_0 =\sqrt {\frac {L^2}{D}(\omega_n+i(\mu_1 -\mu_2)}$. 
In expression Eq.14 the integration contour $C$ runs around zeros of $det(p)$.
Let us note that although Eqs.10,11 are formally divergent, their 
contribution
to $ \langle (\frac {d \Delta \Omega }{d \theta} )^2 \rangle $
is finite.

Let us consider the case $L_s \ll  L, d$ when results 
do not depend on $\omega_s$. 
To define boundary conditions for Eq.13 it is convenient to introduce
operators
\begin{equation}
S_{\pm}=\frac {1}{2}\biggl[1 \pm \biggl(\bbox{m} (z; \theta_1)
\bbox{\sigma_1}\biggr)
\biggl (\bbox{m}(z; \theta_2) \bbox{\sigma_2}\biggr)\biggr] 
\end{equation}
Then the boundary conditions for $\Psi_m (z; \gamma ,\alpha)$ are:
$\frac {d}{dz} S_{+}\Psi_m = 0$ at $z=\pm (d+L/2)$ and $S_{-}\Psi_m = 0$
at $L/2 < |z|< d+ L/2$ .  
As a result, a solution of the eigenvalue problem Eq.13 gives the
following
spectral
determinant
\begin{eqnarray}
det (ip) = \biggl[\sinh p \sinh (1+2d/L)p  + \biggl(1-\cos (\theta_1
-\theta_2)\biggr)\frac 
{1+\cosh (2dp/L)}{4}\biggr]\times \nonumber\\ \biggl[1+\cosh (2+2d/L)p - 
\biggl(1+\cos (\theta_1
+\theta_2)\biggr)
\frac {1+\cosh (2dp/L)}{2} \biggr] \end{eqnarray}
Integrating over the chemical potentials, suming over
Matsubara frequencies and considering case $L_T \equiv \sqrt {D /
T} \gg L; d$ we get
\begin{equation}
G=\frac {2S}{(4\pi L)^2}
(D/L^2)^2 \int \limits_{0}^{\infty } dp p^5 \Phi (p) 
\biggl [ 1-\frac {1}{2} \frac
{d}{d\theta } \Biggl(\frac {\sin 2\theta }{1+\Phi (p) \frac {1-\cos 2\theta
}{4}}\Biggr)\biggr]
\end{equation}
where 
\begin{equation} \Phi (p) = \frac {1+\cosh 2dp/L}{\sinh p \sinh (1+2d/L)p}
\end{equation}

The case when $L_s \gg L$ can be studied in the same way giving Eq.6.

Work of B.S. was supported by Division of Material Sciences, U.S.National
Science Foundation under Contract No.DMR-9625370.

\newpage

\begin{figure}
  \centerline{\epsfxsize=10cm \epsfbox{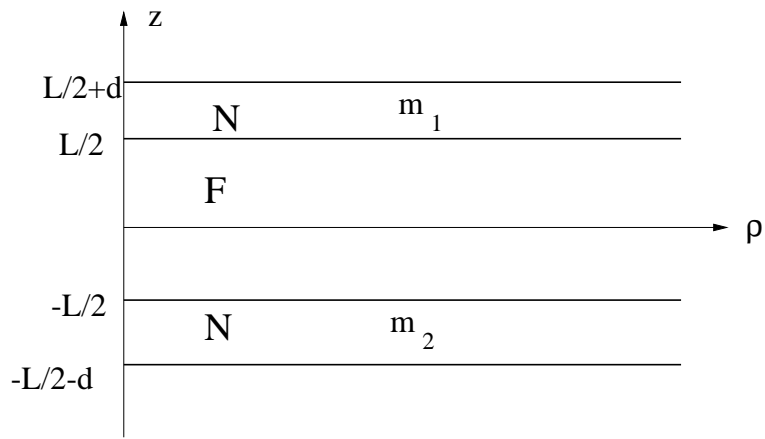}}
  \caption{A schematic picture of the ferromagnet (F)-normal metal (N)-
ferromagnet system} \
  \label{fig:fig1}
\end{figure}

\newpage

\begin{figure}
  \centerline{\epsfxsize=10cm \epsfbox{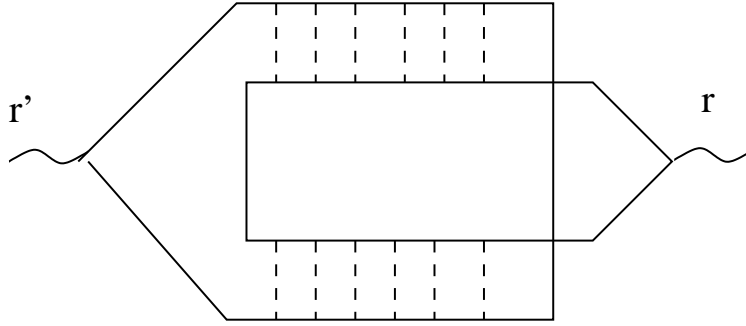}}
  \caption{Diagrams for calculation of the correlation function
$\langle\delta N(\mu_{1},\theta_{1}) \delta N(\mu_{2},\theta_{2})\rangle$.
Solid lines correspond to electron Green
functions and dashed lines correspond to the correlation function of the
scattering potential $<U(\bbox{r})U(\bbox{r}')>$.} \
  \label{fig:fig2}
\end{figure}

\end{document}